# Approach to Dark Spin Cooling in a Diamond Nanocrystal

*Abdelghani Laraoui[*], and Carlos A. Meriles*[1]*

[*]Department of Physics, CUNY-City College of New York, New York, NY 10031, USA

## ABSTRACT

Using a Hartman-Hahn protocol, we demonstrate spin polarization transfer from a single, optically-polarized nitrogen-vacancy (NV) center to the ensemble of paramagnetic defects hosted by an individual diamond nanocrystal. Owing to the strong NV-bath coupling, the transfer takes place on a short, microsecond time scale. Upon fast repetition of the pulse sequence we observe strong polarization transfer blockade, which we interpret as an indication of spin bath cooling. Numerical simulations indicate that the spin bath polarization is non-uniform throughout the nanoparticle averaging approximately 5% over the crystal volume, but reaching up to 25% in the immediate vicinity of the NV. These observations may prove relevant to the planning of future bath-assisted magnetometry tests.

**KEYWORD**S: diamond nanocrystals, nitrogen-vacancy centers, paramagnetic centers, optically detected magnetic resonance, polarization transfer

---

[1] To whom correspondence should be addressed. E-mail: cmeriles@sci.ccny.cuny.edu

Nanocrystalline diamond provides a unique platform material with extreme properties and multifunctionalities enabling an ever-growing range of applications.[1] For example, diamond nanoparticles show promise for the fabrication of photonic nanostructures[2] as well as in energy storage.[3] Nanodiamond is comparatively less toxic to biological processes than other classes of carbon nanoparticles, which makes it adequate for various biomedical applications including drug monitoring,[4] biomolecular labeling,[5] and therapeutics.[6]

Diamond nanocrystals typically host a range of paramagnetic defects, some of which can be exploited for high-resolution magnetometry. A well-known example is the NV center, a spin-1 defect comprising a substitutional nitrogen atom and an adjacent vacancy that can be optically initialized and readout. The NV from a diamond nanocrystal attached to the tip of an atomic force microscope has already been used to image a patterned ferromagnetic surface[7] and similar studies have been reported using an engineered diamond nanostructure.[8] Unfortunately, surface defects and other impurities (mostly nitrogen atoms) typically create a fluctuating magnetic field at the NV site. The consequence is a shortened NV lifetime and, correspondingly, a reduced sensitivity to external magnetic fields, which negatively impacts projected applications such as nanoscale electron and nuclear spin sensing.[9,10]

While most work thus far has focused on isolating the NV center from the environment, it has been pointed out that bath spins are not necessarily detrimental to sensing so long as one can initialize and control their time evolution.[11-13] For example, in a geometry where the NV occupies an inner position within the nanocrystal and paramagnetic centers are predominantly near the surface, bath spins may serve as intermediaries to amplify the effect of processes at or near the nanocrystal boundaries.[11,13] Clever schemes of 'environment-assisted' magnetometry tailored to nanocrystal-hosted NVs have been proposed that promise to attain enhanced sensitivity, particularly when the spin environment can be fully polarized.[12]

Here we present experimental results demonstrating our ability to dynamically transfer NV spin polarization to surrounding paramagnetic centers in an individual diamond nanoparticle. We find that the spin transfer takes place on a fast time scale (~4 μs) in a unidirectional, diffusive fashion, from the NV into the bath, which we attribute to strong intra-bath couplings. We demonstrate that the NV serves not only as the source of spin alignment but also as a sensor to probe the bath polarization, at least in its immediate vicinity. For this purpose, we build on the notion of spin transfer blockade, *i.e.*, the idea that spin transfer from the NV into the bath is partly inhibited when neighboring spins are already polarized. We experimentally test this concept by varying the time interval between consecutive applications of the Hartman-Hahn protocol. Further, we construct a model of spin transfer that explicitly takes into account the sequence repetition rate, and show that, in principle, one can calculate the spatial distribution of bath polarization if the NV dependence on the Hartman-Hahn contact time is known.

**RESULTS AND DISCUSSION**

The physical properties of NV centers relevant to this work are well known[14] and will be reviewed only briefly. The NV has a triplet ground state with a zero-field splitting of 2.87 GHz. Upon green laser illumination, the NV undergoes a transition to an excited triplet state followed by intersystem crossing to a singlet state that preferentially decays into $|m_S = 0\rangle$; almost complete NV polarization occurs after ~1 μs excitation. Intersystem crossing has a lower probability if excitation takes place from $|m_S = 0\rangle$, implying that the fluorescence intensity from short (<1 μs) laser pulses correlates with the spin state predating detection, hence allowing for optical readout of the spin state. We apply an external magnetic field aligned with the NV symmetry axis to separate the $|m_S = \pm 1\rangle$ states, and selectively address one of the two possible transitions (in the present case, the $|m_S = 0\rangle \leftrightarrow |m_S = -1\rangle$ transition). Upon offsetting and normalizing the maximum fluorescence contrast (30%), we obtain an NV signal in the [-1,1] range, where the maximum (minimum) corresponds to $|m_S = 0\rangle$ ($|m_S = -1\rangle$). For the present experiment we use commercial diamond nanoparticles with diameter in the 30-50 nm range. While NVs are scarce in particles of this size,[15] other paramagnetic centers of various types are fairly abundant and form the so-called 'spin bath'. Optical excitation of these defects does not yield fluorescence, the reason why they are typically referred to as 'dark spins' (DS); in the text that follows we use the terms 'dark spin' or 'spin bath' exchangeably.

An initial spectroscopic characterization of the DS system in the nanocrystal we investigate herein is presented in Figs. 1a and 1b: In this case we monitor the NV optical signal upon application of a Hahn-echo sequence of fixed duration ($2\tau_1$=1.2 μs) as we scan the frequency of a second microwave field. The spectrum reveals the presence of substitutional nitrogen (P1-centers, solid line) as well as other paramagnetic centers, most likely surface defects.[11,16,17] The strong hyperfine splittings (86 MHz and 114 MHz) of the P1-centers due to the [14]N nucleus[18] (a spin-1 system) make broadband excitation of the bath technically demanding. Fortunately, [14]N spins are mostly pinned along their local field gradients, hence merely acting as a source of magnetic field heterogeneity with virtually no influence on the electron spin dynamics. Thus, even if at the expense of signal contrast, we can gain meaningful information on the DS ensemble if we selectively excite only one of the DS spectral transitions. A first example is shown in Fig. 1c, where we observe dark spin Rabi oscillations resulting from microwave pulses of variable duration; Fig. 1d shows the DS coherence decay upon application of a Hahn-echo sequence. In both cases (as well as in the experiments we present below), we use microwave excitation at the central dark spin transition (943 MHz). It is worth noting that the Hahn-echo sequence does considerably extend the DS coherence lifetime from the 62 ns observed upon application of an indirectly-detected Ramsey sequence[11] (*i.e.*, a sequence identical to the one sketched in Fig. 1a but without the midpoint DS spin inversion). Because the homo-spin interaction must be immune to π-pulses, we conclude that much of the observed spectral broadening in Fig. 1b is the result of field heterogeneity, perhaps created by slowly reconfiguring charge distributions in the nanoparticle.

To transfer NV polarization to the dark spin ensemble we implement a Hartman-Hahn (HH) protocol.[19] The central idea is that while energy mismatch generally prevents flip-flops between different spin species in the lab frame, polarization can be transferred in the rotating frame when the amplitudes of two simultaneous microwave fields are chosen to produce equal Rabi splittings. In the present case, direct spin flip-flops between the NV and bath spins are possible in the vicinity of 50 mT (where the frequency of the NV $\left| m_S = 0 \right\rangle \rightarrow \left| m_S = -1 \right\rangle$ transition roughly coincides with that of a spin-1/2 paramagnetic center), but is suppressed elsewhere. The Hartman-Hahn protocol is precisely conceived to overcome this limitation: Because the Landé factors of the NV center and most paramagnetic defects approach a value g~2, the optimum HH transfer must take place when both microwave fields have the same amplitude.

The spin transfer sequence we use is presented in Fig. 2a: After optical initialization in the $\left| m_S = 0 \right\rangle$ state, we align the NV along the microwave field MW1 using a π/2-pulse followed by a 90-degree phase-shifted 'spin-locking' pulse. We then apply a second in-phase π/2-pulse that projects the NV onto the laboratory quantization axis for inspection. In the absence of polarization transfer (or NV relaxation), the chosen phase protocol produces the final NV state

$\left| m_S = -1 \right\rangle$. A similar pulse sequence is used for the bath spin *via* a second microwave field (hereafter labeled as MW2) tuned to the DS central transition. The duration of the MW1 and MW2 spin locking pulses can be controlled independently; NV-DS spin flip-flops, however, only take place during the contact time $T_{\text{HH}}$.

Fig. 2b shows the NV response upon application of spin locking pulses of identical, fixed duration (30 μs) as we scan the amplitude of MW2. We find a sensitive dependence on the microwave power, initially growing with increasing spin-lock field to gradually vanish once the amplitude of MW2 exceeds that of MW1 (2.23 G). Physically, we interpret the observed response as a manifestation of the polarization transfer: When the Hartman-Hahn matching condition is met, the NV exchanges energy with the less polarized bath to reach a common spin temperature. From the signal dependence on the duration of the spin-lock pulse (Fig. 2c), we find that the process of thermalization takes place on a time scale $T_{HH}^{opt} \sim 4$ μs. The latter is substantially shorter than the NV $T_{1\rho}$ time — the spin-lattice relaxation time in the rotating frame — observed to be of the order of 100 μs (see below). The fast, monotonic decay found under Hartman-Hahn matching (red circles) points to a unidirectional polarization transfer, from the NV to the bath. This is in contrast with the oscillations typical of cross-polarization experiments between proton and carbon spins observed in NMR,[20,21] and suggests a scenario where strong intra-bath couplings swiftly transport polarization away from the NV *via* spin diffusion.

Assuming a Hartman-Hahn sequence of near optimum duration ($T_R \sim T_{SL} = T_{HH} = T_{HH}^{opt}$), we can get a rough estimate of the total time required to polarize the spin bath by calculating the number of paramagnetic centers in the nanocrystal. For this purpose, we use the measured $T_2$ time (410 ns) to estimate the average separation between dark spins ($d_{DS} \sim 5$ nm), and obtain a paramagnetic center concentration $C' \sim 100$ ppm (where the prime indicates that the we are only considering the fraction of dark spins associated with the central transition, see below). In the limit of a uniform spatial distribution, the latter leads to roughly 200 dark spins for a 30-nm-diameter nanocrystal implying that at least $\sim 1$ ms of multiple NV-pumping/spin-transfer cycles would be required to attain full DS polarization (this time must scaled up by a factor $\sim 2$ if dark spins associated to hyperfine-shifted transitions are included in the calculation). Naturally, this latter scenario rests on bath spins having a sufficiently long $T_{1\rho}$ time, a condition unlikely to be met here: Although a direct determination of the rotating frame lifetime is difficult, we surmise that a value comparable to that of the NV ($\sim 100$ μs) serves as a reasonable guess. In this case, repeated application of the Hartman-Hahn protocol is expected to produce at best a total of $T_{1\rho}/T_{HH}^{opt} \sim 25$ fully polarized spins, and thus one anticipates an average DS polarization on the order of a few percents over the particle volume.

Implicit in the estimates above is the notion that strong intra-bath couplings transport the NV polarization almost instantaneously throughout the nanoparticle. Clearly, this is only a rough approximation, as a finite spin diffusion coefficient will lead to non-uniformity, with bath spins closer to the NV more strongly polarized than the rest. A non-uniform bath polarization impacts the efficiency of the Hartman-Hahn protocol in the sense that polarization transfer is less likely once all neighboring spins have been polarized. Qualitatively, we find signs of this 'spin blockading' process in Fig. 2: In Fig. 2c, for example, we plot the NV response after application of the Hartman-Hahn protocol as we scan the frequency of MW2 in the vicinity of the bath central transition; both the spin locking and contact times are 30 μs. While we observe the typical Lorentzian dependence on the offset, we find that the NV signal after resonant transfer has a different-than-zero, negative value, hence pointing to partial bath polarization.

To more thoroughly understand this observation, we model the spin transfer process *via* the set of equations

$$\frac{dP'_{DS}(\vec{r},t)}{dt} = W(\vec{r})\left(P_{NV}(t) - P'_{DS}(\vec{r},t)\right) - \frac{P'_{DS}(\vec{r},t)}{T_{1\rho}} + D'\nabla^2 P'_{DS}(\vec{r},t),$$  (1)

and

$$\frac{dP_{NV}(t)}{dt} = \frac{8}{V_{uc}}\int_V d^3r\, C'(\vec{r})W(\vec{r})\left(P'_{DS}(\vec{r},t) - P_{NV}(t)\right) - \frac{P_{NV}(t)}{T_{1\rho}} + \delta(t - nT_R)\left(P_{NV}^{(i)} - P_{NV}(t)\right),$$  (2)

where $P_{NV}(t)$ ($P'_{DS}(\vec{r},t)$) represents the nitrogen-vacancy (dark-spin) polarization at (position $\vec{r}$ and) time $t$, $W(\vec{r})$ is the polarization transfer rate in the rotating frame between the NV and a given dark spin at $\vec{r}$. In Eq. (2), the last term takes into account the periodic NV re-initialization into a state $P_{NV}^{(i)}$ after a time $T_R$, $V_{uc}$ is the volume of the unit cell in diamond, $V$ denotes the nanocrystal volume, and $C'(\vec{r})$ is the probability of finding a dark-spin at position $\vec{r}$; finally, $D' = d'^2_{DS}/\left(50\, T_{2DS}^*\right)$ is the diffusion constant for a bath of average inter-spin distance $d'_{DS}$.[19] Throughout Eqs. (1) and (2) we use a prime to highlight the fact that the paramagnetic centers exchanging polarization with the NV are only those within the subset associated with the central transition in the hyperfine-split EPR spectrum (Fig. 1b). For our present experimental conditions, the latter amounts to approximately half the total number of dark spins.

Finding a numerical solution to the set of coupled equations above is not simple in the general case, but we can gain relevant insight if we first think of a NV spin equally coupled to $N \gg 1$ spin neighbors in the limit of zero spin diffusion. Denoting with $W$ the transfer rate between the NV and an individual spin in the bath, the system reaches a common spin temperature after a time interval $1/(NW)$. Because in a typical experiment several repeats are necessary to reach an acceptable signal-to-noise ratio, the time $T_R$ separating two consecutive applications of the Hartman-Hahn sequence is important in defining the final NV polarization $P_{NV}^{(f)}$ (and thus the system spin temperature at equilibrium). At short times $t$ (*i.e.*, when the accumulated number of HH repeats $n$ is small, see Eq. (2)) the common temperature should approach infinity because the bath is initially unpolarized. Successive repetitions of the protocol, however, make the bath polarization build up, thus leading to a change in $P_{NV}^{(f)}$. The exact value — unchanged once a stationary regime has been reached — emerges from the interplay between $T_R$ and the spin-lattice relaxation time $T_{1\rho}$.

We note that, strictly speaking, this 'stationary' regime must be dynamic in the sense that rather than taking time-independent values, the NV and DS polarizations evolve during a given interval $T_R$. In other words, in the limit of large $n$ both $P_{NV}(t)$ and $P_{DS}^{'}(t)$ become periodic functions of time. Fig. 3a shows schematically the anticipated system dynamics for the idealized scenario where NV-DS spin flips are allowed at all times (*i.e.*, where NV and DS preparation is carried out in negligible time and both microwave fields are continuously on): After each NV initialization (signaled by the periodic discontinuities characterizing the green solid line in Fig. 3a) $P_{NV}(t)$ decays quickly to reach a common value with the bath. The decay is bi-exponential: The short-time scale is controlled by $1/(NW)$ (the time constant describing the NV-DS coupling) while the long-term decay results from spin lattice relaxation (here assumed to be the same for both the NV and DS systems). A stationary dynamic equilibrium is reached when the dark spin states before and after application of the Hartman-Hahn protocol are the same.

Experimentally, we expose the influence of the repetition rate on $P_{NV}^{(f)}$ in Fig. 3b: Here we apply the pulse sequence of Fig. 2a using a fixed spin locking time of 30 μs, sufficiently long for the NV-DS system to thermalize. Red squares reproduce the data points of Fig. 2b with consecutive HH contacts following almost immediately one after the other; blue triangles are the result of a similar experiment but this time a delay of 100 μs is introduced between the end of a repeat and the beginning of the next one. Both curves are described well by single exponentials but the time constant characterizing the red curve (4.6 μs) is longer than that corresponding to the blue curve (3.7 μs) in qualitative agreement with the idea that the spin transfer is faster when neighboring spins are less polarized. Something similar can be said about the clear difference in the NV signal amplitude $P_{NV}^{(f)}$ following the HH contact: By increasing the time separation between consecutive repeats, spin-lattice relaxation is more effective in annihilating the system polarization after the spin transfer. As a result, subsequent application of the Hartman-Hahn protocol exposes the NV to a less polarized bath, hence bringing the value of $P_{NV}^{(f)}$ closer to zero (blue solid line).

At first sight it may seem strange to find in Fig. 3b a single-exponential dependence rather than the bi-exponential decay sketched in Fig. 3a, especially given the moderate $T_{1\rho}$ time (~100 μs). This behavior, however, is a direct consequence of the chosen pulse protocol: In a sequence where the duration of the NV spin locking pulse is kept constant (as in the experiments of Figs. 2b and 3b), spin lattice relaxation acts equally on each data point allowing us to conveniently separate the effect of the bath on the NV upon thermal contact. The price to pay is a less-than-optimum fluorescence contrast: From Fig. 3b we find that as the contact time approaches zero, the NV signal tends to a value different from −1 (otherwise the expected response in a regime of negligible spin-lattice relaxation).

Naturally, one can explicitly reintroduce the effect of $T_{1\rho}$ in the NV signal if, rather than setting $T_{SL}$ to a fixed value, the NV spin locking time is systematically chosen to coincide with the contact time (*i.e.*, MW1 and MW2 are on for the same time). The result of using this alternate Hartman-Hahn sequence is shown in Fig. 3c: In the two cases where thermal contact is allowed (red squares and blue triangles), the NV signal now displays the expected bi-exponential behavior with a slow time constant defined by the value of $T_{1\rho}$ (as obtained from the NV decay in the absence of spin transfer, black circles). Echoing the observations of Fig. 3b, the NV signal corresponding to the point where the transition from fast to slow decay takes place can be altered by introducing an additional 100 μs delay between consecutive repeats (blue triangles).

Our ability to influence the Hartman-Hahn signal through the repetition rate has important practical implications because it gives us the opportunity to indirectly monitor the spin polarization in the NV vicinity (otherwise a difficult task when probing a nanoscale dark spin ensemble). In particular, it should be possible to estimate $P'_{DS}(\vec{r},t)$ using $P_{NV}(t)$ as a known input in Eqs. (1) and (2). For our present purposes, however, we can greatly simplify the computation if we focus on the time averages rather than on the instantaneous polarizations. To this end, we take the time integral over a HH cycle and recast (1) and (2) as

$$0 = \langle P_{NV} \rangle W(r) - \langle P'_{DS}(r) \rangle \left( W(r) + \frac{1}{T_{1\rho}^{DS}} \right) + \frac{D'}{r^2} \frac{\partial}{\partial r} \left( r^2 \frac{\partial \langle P'_{DS}(r) \rangle}{\partial r} \right), \qquad (3)$$

and

$$\frac{\Delta P_{NV}}{T_R} = \frac{8}{V_{uc}} \int_V d^3r \, C'(r) \, W(r) \left( \langle P'_{DS}(r) \rangle - \langle P_{NV} \rangle \right) - \frac{\langle P_{NV} \rangle}{T_{1\rho}}, \qquad (4)$$

where brackets indicate average over the interval $T_R$, and $\Delta P_{NV} \equiv P_{NV}^{(f)} - P_{NV}^{(i)}$. In (3) and (4) we use the fact that the average of the derivative of a periodic function cancels, and we assume for simplicity radial symmetry around the NV.

Rather than solving the coupled set of equations, here we follow a simpler approach that relies on the measured value of $\langle P_{NV} \rangle$ to determine $\langle P'_{DS}(r) \rangle$ *via* Eq. (3). Eq (4) then allows us to check for self-consistency by comparing the calculated and measured values of $\Delta P_{NV}$. In our calculations, we model the polarization transfer rate as $W(r) = W_0 (r_{min}/r)^6$, where $r_{min} \sim 4.5$ nm indicates the distance to the closest dark spin and $W_0$ is the rate of spin transfer between magnetic dipoles separated a distance $r_{min}$.

Fig. 3d shows the calculated average polarization in the spin bath as a function of the distance to the NV for the case $T_R \sim T_{HH}$ in the region $r \geq r_{min}$ (where the dark spin concentration is different from zero). The solid line is the result of solving Eq. (3) using $D' \cong 10^{-12}$ m$^2$/s, while the dotted line corresponds to the same calculation but assuming negligible spin diffusion (*i.e.*, $D' \sim 0$). For the first case, we find that $\langle P'_{DS}(r) \rangle$ reaches up to $\sim$50% in the immediate vicinity of the NV (corresponding to an average dark spin polarization of 25% if the full spin bath is considered) and slowly decays at longer distances (without completely vanishing, however, within the particle volume). The opposite happens when spin diffusion is quenched, leading to significantly enhanced dark spin polarization near the NV. Note that although one can hardly imagine accelerating spin diffusion in the bath without altering the dark spin concentration (and thus the system coherence and spin-lattice relaxation times), it should be possible to effectively bring down the value of $D'$ by articulating the Hartman-Hahn protocol with homo-spin decoupling schemes.

**CONCLUSION**

Complementing prior strategies conceived to transfer polarization from an NV to individual spins in its immediate vicinity, the Hartman-Hahn sequence offers a versatile, alternate route in situations where the target spin system forms a strongly self-interacting bath. Our experiments provide an initial demonstration for diamond nanocrystals, where, when present, the NV is typically surrounded by multiple paramagnetic defects of diverse kinds and coupling constants. The polarization attained in the bath after several repeats ultimately impacts the NV response; this back-action on the NV spin dynamics can be controlled by altering the Hartman-Hahn repetition rate, hence providing a handle to probe the bath polarization. The latter makes the present approach of interest in various applications where attaining the highest spin bath order is important, particularly in mesoscale spin ensembles, difficult (if not impossible) to probe using inductive magnetic resonance.

Our modeling suggests that spin diffusion may play an important role, altering the spatial distribution of polarization from what would otherwise result in the absence of intra-bath couplings. In a geometry where the NV occupies a more secluded, inner region of the nanoparticle the dark spin concentration seems sufficient to transport polarization over tens of nanometers, limited by the finite particle size. Assuming sufficiently long spin-lattice relaxation times (*e.g.*, possible at low temperatures), this restricted diffusion may prove instrumental in attaining near full spin bath polarization throughout the nanoparticle. Alternatively, future experiments may be designed to partially hinder intra-bath interactions so as to concentrate the bath polarization to within the immediate NV vicinity. We note that one regime or the other may prove useful under different conditions: For example, partially polarized spins nearer to the particle surface than the NV could serve as intermediaries to amplify the signal generated by target magnetic sources (such as paramagnetic molecules or magnetic nanoparticles) attached to the diamond nanocrystal.[13] Conversely, the ability to more strongly polarize neighboring spins could prove useful to protect the bath polarization from enhanced relaxation near the nanocrystal walls, or for the implementation of 'bath-assisted' magnetometry schemes[12] conceived to enhance the detection sensitivity of non-local magnetic fields.

Besides polarizing electronic spins, the Hartman-Hahn protocol can be adapted to transfer polarization to nuclear spins as well. Given the large differences between the electronic and nuclear gyromagnetic ratios (typically exceeding three orders of magnitude), energy matching in the rotating frame is impractical. This problem, however, can be circumvented by a 'hybrid', laboratory/rotating frame matching, in which the amplitude of the electron spin-locking microwave is chosen so as to produce an energy splitting coincident with that induced on the nuclear spins by the static magnetic field. Such strategy has already been used extensively to dynamically polarize $^{13}$C spins in bulk diamond with large nitrogen content,[22] and, more recently, in high purity crystals using optically pumped NVs.[23] Along the lines of the experiments presented here, an intriguing possibility is the investigation of the polarization transfer near 50 mT, where level anti-crossing in the excited NV state can strongly influence the polarization of the host nitrogen nuclear spin[24] and neighboring carbons.[25] Alternatively, the principles of Hartman-Hahn transfer could be extended to polarize nuclei other than $^{13}$C, a tantalizing possibility made all the more plausible by the recent demonstration of proton spin detection using shallow NVs.[26,27]

On a final note, we mention that during the writing of this manuscript we learned that experiments similar to those reported herein have been conducted at Harvard.[28]

**METHODS**

In our experiments we use commercial, HPHT-grown nanocrystals (MSY 0-0.1 GAF

Microdiamant), separated from the original solution after centrifuging and re-dissolving several times. After drop casting on a glass substrate, we attain a homogeneous collection of isolated nanoparticles with diameter in the 30-50 nm range. We use the green light from a laser (Coherent Compass, 532 nm) and a high-numerical-aperture objective (Nikon CFI Plan Fluor 100x, NA = 1.35) from a purpose-built confocal microscope to excite NV fluorescence. Light collection is carried out by the same objective and subsequently filtered for wavelengths above 630 nm by a dichroic beam splitter and a long pass filter. For detection, we use a single mode fiber (4 μm in diameter) with a 50% beam splitter (2 outputs) connected to 2 avalanche photodiodes (SPQR-14 Perkin-Elmer).

As described in Ref. [11], the sample sits on a stage between the objective of the confocal microscope described above and the tip of a commercial atomic force scanner (5500 Agilent AFM/SPM). This configuration allows us to image the sample topography (*i.e.*, identify the nanoparticle positions on the glass substrate) and subsequently collect NV fluorescence from a selected diamond crystal. For the system of nanoparticles we investigate here, we find that a small fraction (1 %) hosts NVs, which, in turn, typically prove to be single NV centers as determined by photon-antibunching experiments. The diamond crystal chosen for the present experiments has an approximate diameter of 30 nm.

We manipulate the NV and dark spins *via* the microwave fields from a 25 μm copper wire spanning the surface of the glass substrate. Microwave pulses are produced by two signal generators (Rohde&Schwartz SMB100A 9 kHz-6 GHz) connected to two fast switches (Mini-Circuits ZYSWA-2-50DR, rise time of 6 ns). The microwave signals are combined using a power splitter and sent through a high power (30 W) amplifier (Amplifier Research 30S1G4, 0.8-4.2 GHz); the typical duration of a $\pi/2$-pulse is ~20 ns. We control the microwave phase *via* a digital phase shifter (MaCom). All experiments are carried out in the presence of an external magnetic field (~337 G) whose direction is chosen to coincide with the NV axis, and we selectively manipulate the $|0\rangle \rightarrow |-1\rangle$ transition. Upon implementing a Hahn-echo (inversion-recovery) protocol, we find that the typical NV $T_2$ ($T_1$) time in these nanocrystals is ~3 µs (0.5 ms).

*Conflict of Interest:* The authors declare no competing financial interest.

*Acknowledgement:* We thank Jonathan Hodges for assistance with our experimental setup as well as Sen Yang and Ya Wang for helpful discussions. We acknowledge support from the National Science Foundation through Grants NSF-1111410 and NSF-0545461.

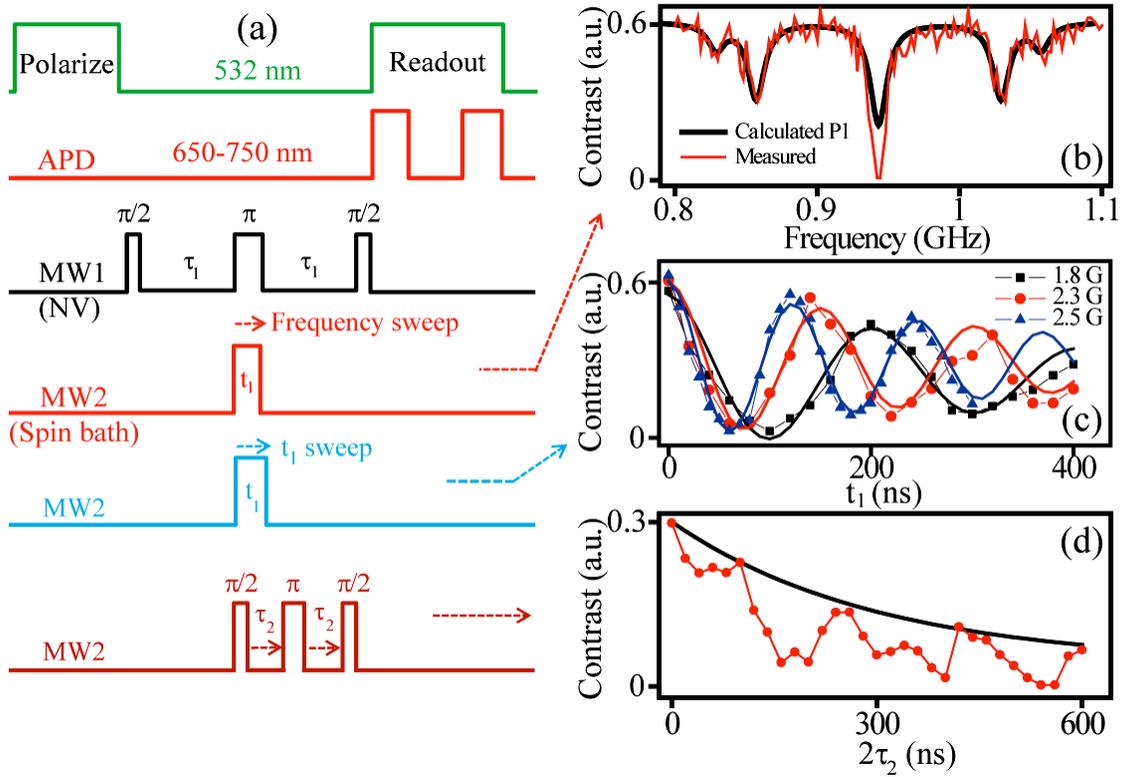

**Figure 1.** (a) Schematics of alternate double resonance protocols. In all cases, we monitor the NV spin echo signal for a fixed interpulse interval $\tau_1$. (b) Observed response as one scans the frequency of MW2; the pulse duration $t_1$ is 80 ns. The black solid line represents the calculated spectrum for nitrogen impurities (P1 centers) in the case of a magnetic field aligned with the [111] axis using the hyperfine couplings from Ref. [18]. (c) Rabi oscillations of the spin bath for three different microwave fields: 1.8 G (dark squares), 2.3 G (red circles), and 2.5 G (blue triangles). The solid lines correspond to exponential fits with a time constant of 400 ns. (d) Spin echo protocol on the spin bath. The solid line is a single exponential decay with time constant 0.41 μs. In all experiments $\tau_1$=0.6 μs, $B_0$ = 336.8 G, and MW1 has a frequency of 1.927 GHz and amplitude of 3 G. In (c) and (d), MW2 has the frequency of 943 MHz, coincident with the bath central transition.

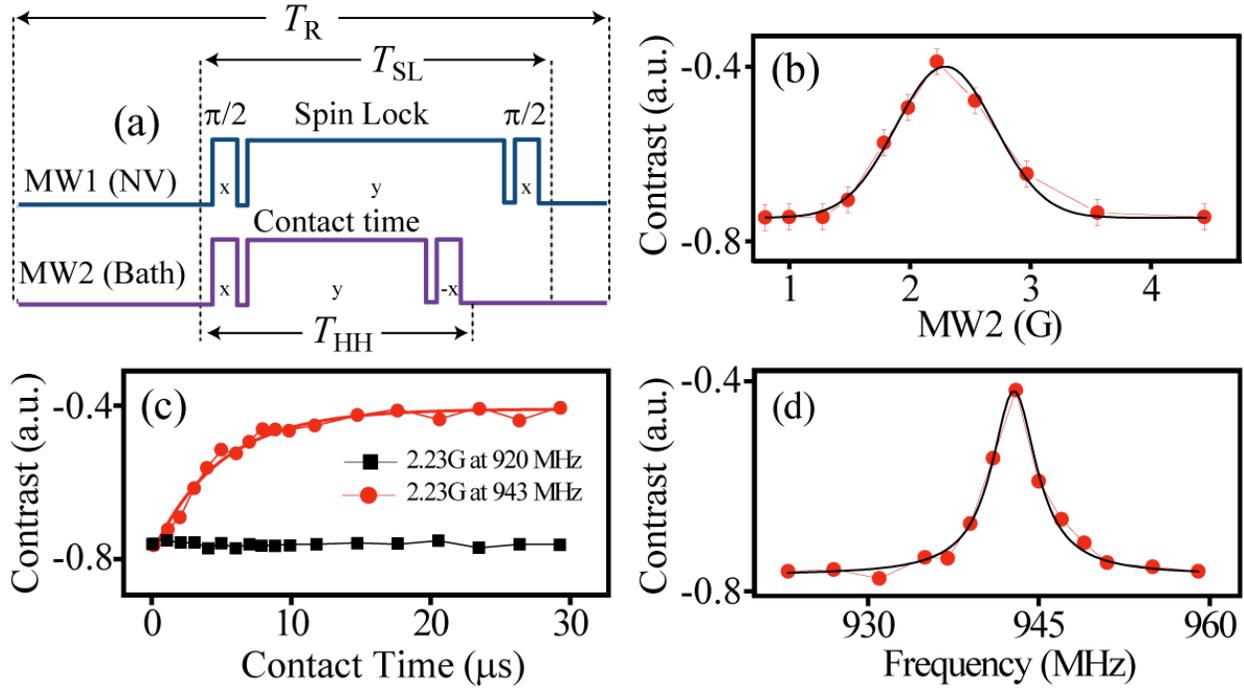

**Figure 2.** (a) Polarization transfer protocol. NV initialization and readout (not shown) are carried out as in Fig. 1. (b) NV signal for a spin locking field MW2 of variable amplitude at 943MHz. (c) NV response as a function of the contact time interval; MW2 has an amplitude of 2.23 G and a frequency of either 943 MHz (red circles) or 920 MHz (black squares). The solid red line corresponds to a single exponential decay with a time constant of 4.6 μs. (d) Same as in (b) but for a MW2 spin-locking pulse of variable frequency and amplitude of 2.23 G. In (b) and (d) the spin-locking and contact times are 30 μs long. In all experiments, MW1 has an amplitude of 2.23 G; all other conditions as in Fig. 1.

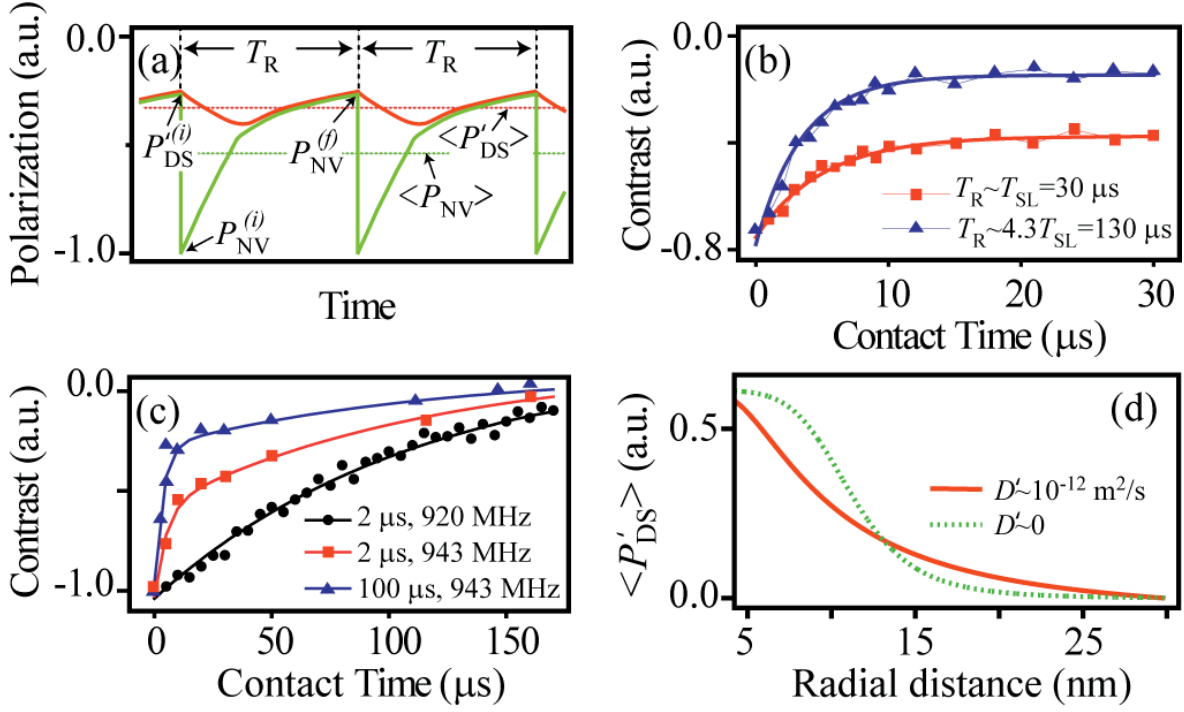

**Figure 3.** (a) Schematics of the NV and dark spin polarizations (green and red solid lines, respectively) as a function of time in the 'quasi equilibrium' regime. (b) NV response as a function of the contact time for a fixed spin locking time of 30 μs. Red squares (blue triangles) correspond to the case $T_R$~30 μs ($T_R$~130 μs). (c) Here the NV spin-locking time matches the contact time. The frequency of MW2 is either 943 MHz (red squares and blue triangles) or 920 MHz (black circles). The delay between successive applications of the Hartman-Hahn protocol takes the value 2 μs (red squares and black circles) or 100 μs (blue triangles). The red (blue) solid lines are bi-exponential fits with time constants 4.6 μs (3.7 μs) and 96 μs (97 μs). The solid black line corresponds to a single exponential fit with time constant 96 μs. (d) Calculated magnitude of the spin bath polarization (dark spins associated to the central transition only) as a function of the radial distance to the NV (Eq. (3)) with or without spin diffusion (red solid line and green dotted line, respectively).